\documentclass[aip,preprint]{revtex4-1}
\usepackage{natbib}
\usepackage{graphicx}
\usepackage{rotating}
\usepackage{wrapfig}
\usepackage{amsmath}
\usepackage{bm}

\begin{document}


\title{Symmetry breaking in single crystal SrTiO$_3$ plates: EPR manifestations}

\author{B.~F.~Gabbasov}
\email[]{bulgabbasov@gmail.com}
\affiliation{Kazan Federal University, Kremlyovskaya str. 18, 420008 Kazan, Russia}

\author{I.~N.~Gracheva}
\affiliation{Kazan Federal University, Kremlyovskaya str. 18, 420008 Kazan, Russia}

\author{A.~A.~Rodionov}
\affiliation{Kazan Federal University, Kremlyovskaya str. 18, 420008 Kazan, Russia}

\author{S.~I.~Nikitin}
\affiliation{Kazan Federal University, Kremlyovskaya str. 18, 420008 Kazan, Russia}

\author{D.~G.~Zverev}
\affiliation{Kazan Federal University, Kremlyovskaya str. 18, 420008 Kazan, Russia}

\author{V.~A.~Trepakov}
\affiliation{Ioffe Physical-Technical Institute, Russian Academy of Sciences, 194021 St.~Petersburg, Russia}

\author{A.~Dejneka}
\affiliation{Institute of Physics, ASCR, v.v.i. Na Slovance 2, 18221 Prague, Czech Republic}

\author{L.~Jastrabik}
\affiliation{Institute of Physics, ASCR, v.v.i. Na Slovance 2, 18221 Prague, Czech Republic}

\author{R.~V.~Yusupov}
\email[]{Roman.Yusupov@kpfu.ru}
\affiliation{Kazan Federal University, Kremlyovskaya str. 18, 420008 Kazan, Russia}


\date{\today}

\begin{abstract}
Electron paramagnetic resonance studies of the impurity Fe$^{3+}$ and Mn$^{4+}$ centers indicate that the symmetry of the crystal structure of millimeter-scale strontium titanate plates above $105$~K lowers in the bulk from cubic to tetragonal, different from that of the conventional low-temperature antiferrodistortive phase. It is shown that the effect does not originate from the residual stress; its magnitude (observed tetragonal distortion) and particular manifestations depend on plate orientation, surface quality and the geometry of a sample. Peculiarities of the observed phenomenon are presented and possible scenario of its realization is discussed.	
\end{abstract}


\maketitle 

\section{Introduction}
Strontium titanate SrTiO$_3$ (STO) is a well-studied model representative of highly-polarizable ABO$_3$ oxides \cite{muller1979srti,barrett1952dielectric,muller1958paramagnetische,fleury1968soft,courtens1972birefringence,cowley1996phase,yamanaka2000evidence,Kvyat2001}. Having cubic perovskite-type structure at room temperature, it obeys a model second-order cubic-to-tetragonal $O^1_h \rightarrow D^{18}_{4h}$   antiferrodistorsive (AFD) phase transition at $T_c \sim 105$~K. Due to its unique properties and ongoing discoveries of new phenomena, STO stays the subject of intense investigations. A variety of beneficial effects arising in STO with doping is especially attractive: it can be endowed with the properties of a ferroelectric \cite{Lemanov97_STO_Pb}, ferromagnet \cite{Taniuchi2016}, conductor \cite{Tufte67} and superconductor \cite{schooley1964superconductivity,schooley1965dependence}, exhibit memristor effect \cite{karg2006electrical,szot2006switching}. This leads to the fact that STO is considered nowadays a material of a choice in a wide range of applications. Further improvement of its performance demands a detailed knowledge of the prevailing defects structure and its impact on the macroscopic material properties. 

Strontium titanate has been well studied in the macroscopic crystalline form. The modifications of the STO properties in the conditions of a restricted dimensionality are investigated in nanoparticles and thin films \cite{Makarova2010,Tyunina2009}. Surface \cite{alexandrov2009first} and interface \cite{oja2012d} properties and the related phenomena are also intensively studied. Oriented single-crystal plates of SrTiO$_3$ are widely used as the substrates for thin films and heterostructures in oxide-based electronics. Therefore, the effects that either occur at the surface of the oriented STO plates or are induced by a surface become important. However, the problem of a mutual influence of the surface and the bulk of crystalline STO has not been addressed thoroughly enough; the extent to which the properties of the bulk are determined by the surface remains unclear. Here, we show that in small millimeter-scale crystalline STO samples this interrelation is quite pronounced and well defined.

Electron paramagnetic resonance (EPR) of the impurity centers that disturb negligibly doped material (paramagnetic probes) is a powerful experimental method for studying a structure of a compound and phase transformations in it. Precisely, the analysis of the EPR spectra modifications of the impurity Fe$^{3+}$  and Gd$^{3+}$ ions~\cite{unoki1967electron,rimai1962electron} allowed for the first time to determine the symmetry of the low-temperature AFD-phase of STO despite the smallness of the distortion magnitude.

In EPR studies, a surface quality and a shape of millimeter-sized samples are usually not considered, as it is generally accepted that the volume of the near-surface region perturbed by the interface is much smaller than the total volume of a sample. Under an assumption that the impurities of the interest are uniformly distributed over a sample volume, EPR spectra should be dominated by the signals from the doped centers located in the unperturbed matrix. However, during our investigations of the high-quality Verneuil grown SrTiO$_3$:Mn and SrTiO$_3$:Fe crystals, it was found that this conventional preposition systematically failed. In particular, the structure of the EPR spectra of well-studied cubic Mn$^{4+}$ and Fe$^{3+}$ centers at $T > 105$~K was found dependent on sample geometry, orientation of its faces relative to crystallographic directions, and face quality. Impurity Mn$^{4+}$ and Fe$^{3+}$ ions are good complementary paramagnetic probes for tracking modifications of STO crystal structure. Both ions form the impurity centers of the cubic symmetry, and even a tiny distortion of the nearest surrounding manifests itself in noticeable change of the EPR spectra.

This paper presents a detailed study of the symmetry breaking that takes place in single-crystal STO samples, predominantly with a shape of a thin though macroscopic ($0.1 \div 1 $~mm thick) plate, using the EPR spectroscopy of the impurity Mn$^{4+}$ and Fe$^{3+}$ ions. It is shown that at $T > T_c$, the symmetry of the (001)- and (110)-oriented plates with imperfect faces, in the bulk, is lowered to tetragonal with the structure different from the AFD-phase. We show also that the effect itself does not originate from the residual stress. Occurrence of the effect and its magnitude are determined by sample geometry, size, and quality of its faces. Symmetry breaking occurs at the temperatures well above the $T_c$ and is related to particular compliance of the STO crystal structure along the $\left\langle 100 \right\rangle$-directions.

\section{Samples and experimental details}
Samples were cut from the high-quality single crystals of STO:Mn grown at Furuuchi Chemical Corp. (Tokyo, Japan) and STO:Fe were provided by Prof. S. Kapphan (Osnabrueck, Germany). Samples were orientated using the Bruker D8 Advance X-ray diffractometer equipped with an Euler cradle with an accuracy of $\pm 1$ degree. All the investigated samples had a right square prism shape with the base side \textit{a} and height \textit{h}. The two limit cases were thin plates with $a \gg h$ (majority of the samples) and bars with $a \ll h$ . The quality of sample faces had two grades: coarsely ground (CG) on a diamond faceplate with a roughness of $\sim 150$ nm and finely polished (FP) with a roughness of $\sim 3$ nm. The roughness was determined as the width of the Gaussian distribution of heights about the mean value according to atomic force microscopy (AFM, Bruker Dimension Fastscan) over the area of $50 \times 50 ~\mu$m$^2$  for the CG-faces and $3 \times 3 ~\mu$m$^2$  for the FP-ones.  

The EPR spectra were measured with a Bruker ESP300 commercial X-band spectrometer. The temperature of the samples in the range of $90 – 300$ K was varied using a liquid nitrogen-flow Oxford Instruments ER4111VT cryogenic system. High Q-factor double-mode ER4116DM rectangular cavity operating in the TE$_{102}$ mode was used.

The modeling and the analysis of the EPR spectra were performed with the use of the Easyspin module \cite{stoll2006easyspin} for Matlab. In the case of a small tetragonal distortion with predominant cubic symmetry, the spin Hamiltonian of a center has the form \cite{abraham1970electron}
\begin{equation}\label{SHam}
\hat{H} = g \beta \mathbf{B} \cdot \hat{\mathbf{S}} + A \hat{\mathbf{S}} \cdot \hat{\mathbf{I}} + B_4\left( \hat{O}^0_4 + 5 \hat{O}^4_4 \right) + B^0_2 \hat{O}^0_2 + B^0_4 \hat{O}^0_4,
\end{equation}
where the first term describes the Zeeman effect, the second corresponds to the hyperfine interaction, the rest describe the “fine” structure of the spectra due to the anisotropy of the crystal field (the third term represents the cubic symmetry term, the fourth and the fifth – the tetragonal one); \textit{g} is the Lande \textit{g}-factor, $\beta$ is the Bohr magneton, $\hat{O_k^q}$ are equivalent Stevens operators. For the Mn$^{4+}$ ion (electron spin $S = 3/2$, nuclear spin $I = 5/2$) the terms containing  $k > 2$ operators are absent, and for the Fe$^{3+}$ ion ($S = 5/2$, $I = 0$), hyperfine interaction term is absent in Eq.~\ref{SHam}. In practice, the  $B_4^0$ parameter values obtained from the fits of the orientation dependences of the EPR resonance fields of the Fe$^{3+}$ ions turned out to be zero within the fit accuracy. In this case, the  $B_2^0$  value is a good measure of the axial component of the crystal field. We note also that the anisotropies of the \textit{g}-factor and the hyperfine interaction constant \textit{A} are neglected due to the small deviation of the symmetry from the cubic one.

\section{Results and discussion}

\begin{figure*}[ht]
	\centering
	\includegraphics[clip=true,width=12cm,angle=-90]{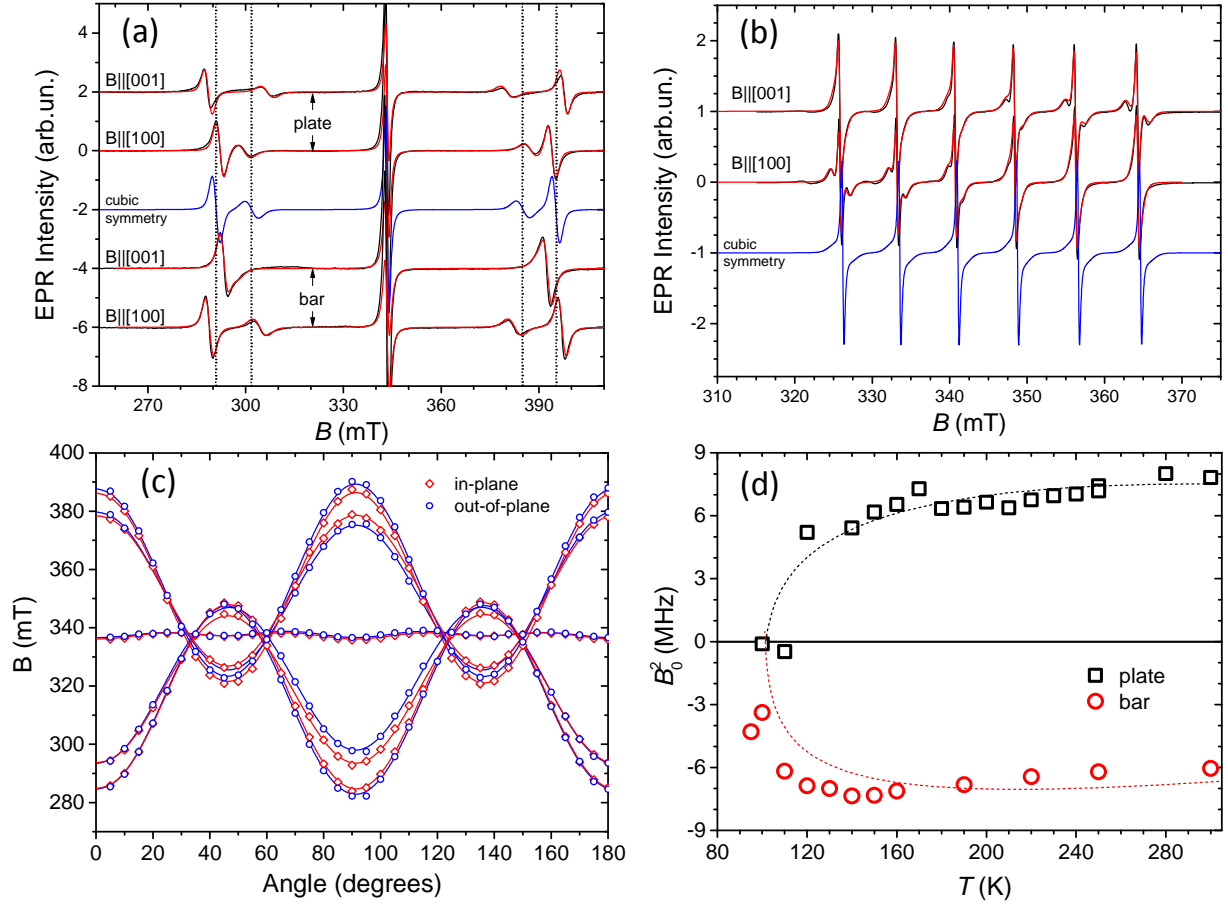}
	\caption{\label{Fig1} EPR spectra of the (001)-oriented single-crystal samples having the shape of a bar and a plate of STO:Fe$^{3+}$ at $T = 300$~K (a) and of a plate of STO:Mn$^{4+}$ at $T = 150$~K (b). The orientations of the samples with respect to the magnetic field \textit{B} are marked in the figure. Black lines are the measured spectra, red ones are the fits and blue lines show the simulated spectra of the non-perturbed cubic-symmetry centers (see text). Orientation dependence of the resonance fields for the Fe$^{3+}$ centers in the STO:Fe (001)-plate (c) with the magnetic field rotated in the (001) and (100) crystallographic planes (rhombs and circles, respectively); its fit using the Hamiltonian (\ref{SHam}) is shown by solid lines. Temperature dependences of the   parameter corresponding to the axial component of the crystal field for a plate (squares) and a bar (circles) shaped samples (d); dotted lines are the guides for an eye.}	
\end{figure*}

Figure~\ref{Fig1},~a and b, shows the EPR spectra of the well-known cubic-symmetry centers of the Fe$^{3+}$ and Mn$^{4+}$ impurity ions in single-crystal STO samples. The samples had a square plate shape with a size of $2\times2\times0.5$~mm$^3$, with the larger faces orientation of (001) and side faces of (100) and (010), respectively. For the SrTiO$_3$:Fe$^{3+}$ crystal, EPR spectra of the sample with a shape of a bar with a size of $0.40\times0.40\times5.0$~mm$^3$ are presented also. All faces of these samples were of the CG-grade. The spectra were measured at temperatures corresponding to the cubic phase of STO that significantly exceeded the $T_c$ (300~K for STO:Fe and 150~K for STO:Mn). For both the Fe$^{3+}$ and Mn$^{4+}$ ions, it is clearly seen that \textit{the spectra measured in the crystallographically equivalent orientations} [001] and [100] \textit{differ from each other}; the spectrum in the [010] direction is identical to the [100]-one (not shown in order to avoid the figure complication).

EPR spectra simulations based on the spin Hamiltonian (\ref{SHam}) are shown in Fig.~\ref{Fig1},~a and b, by red lines. In order to describe the observed differences between the spectra measured with B$\mathbf \parallel \left[ 001 \right]$ and B$\mathbf \parallel \left[ 100 \right]$, a non-zero $B_2^0$-parameter had to be introduced. Its value allowing for satisfactory Fe$^{3+}$ ion EPR spectra simulations in the plate- and the bar-shaped STO samples differed in sign and had the magnitudes of $+ 11$~MHz and $- 14$~MHz, respectively. For the Mn$^{4+}$ ions in the STO plate, the value of the $B_2^0$-parameter 
was $+ 8.5$~MHz. Orientation dependences of the EPR resonance fields of the Fe$^{3+}$ centers in the (001)-oriented STO plate with the magnetic field \textit{B} rotated in the (001) and (100) planes are shown in Fig.~\ref{Fig1},~c. Clearly, the spectrum structure is almost identical at angle values of 0 and 180 degrees (B$\mathbf\parallel[100]$) and reveals a notable difference at 90-degree orientations (B$\mathbf\parallel[010]$ and B$\mathbf\parallel[001]$ for (001)-and (100)-rotations, respectively). Small visible difference in the spectrum structures in these planes at 0 and 180 degrees is due to a small $\sim 4$-degree misalignment of the rotation plane. The solid lines show the fits of the data using spin Hamiltonian (\ref{SHam}). Here, the axial crystal field component $B_2^0 = +11$~MHz had to be taken into account. In all cases, for Fe$^{3+}$ center, other parameter values were $g = 1.999$, $B_4 = 4.9$~MHz; for Mn$^{4+}$, $g = 1.994$, $A = - 243$~MHz. These values are in full accordance with the known data \cite{muller1958paramagnetische,muller1979srti} and were used for simulation of the unperturbed cubic Fe$^{3+}$ center spectrum in Fig.~\ref{Fig1},~a. Experimental EPR spectra of unperturbed cubic-symmetry Fe$^{3+}$ centers in the cube-shaped samples cut from the same boule are shown below. Thus, for two transition metal ions, Fe$^{3+}$ and Mn$^{4+}$, with different electronic configurations, both having an orbital singlet ground state, a symmetry lowering of the impurity centers to tetragonal is observed in thin plates and bars. This clearly indicates the anisotropy of the crystal field in the Ti-site that originates from the STO matrix and is not related to the properties of the impurity centers themselves. The only source of anisotropy for the studied samples was their shape, namely, the difference in size along [001] direction, on one hand, and [010] and [100], on the other. Impurity centers of Fe$^{3+}$ ions due to a full and clear splitting of the EPR spectrum into five fine-structure components are evidently more suitable for studies of the crystal field variations than Mn$^{4+}$; therefore, all further results were obtained on the SrTiO$_3$:Fe crystals. More details on the results obtained on STO:Mn samples can be found in Ref.~\onlinecite{gabbasov2018experimental}.

Temperature dependence of the $B_2^0$  parameter for the STO:Fe plate- and bar-shaped samples is presented in Fig.~\ref{Fig1},~d. For both samples, tetragonal lattice distortion is found in the range from 300 K down to the $T_c \sim 105$~K. For a plate, $B_2^0$ value turns to zero; on further cooling, typical EPR spectra for the AFD-state with a set of symmetry-related centers are observed \cite{unoki1967electron}. For a bar, $B_2^0$ value did not reach zero value but suddenly switched nearby the $T_c$ from a simple structure originating from a single center to a number the symmetry-related ones. Note that indicated $B_2^0$ values are of the same order as the maximum $B_2^0$ value of 11.2~MHz found for the Fe$^{3+}$ centers in STO at low temperatures in the AFD-phase. Thus, the magnitude of the distortion in thin plates and bars is far from negligible. Also, the signs of the $B_2^0$ parameters in combination with the signs of other parameters of the spin-Hamiltonian (\ref{SHam}) determined in Ref.~\onlinecite{dobrov1959electron} indicate that the crystal structure of the (001)-plates is compressed and that of the [001]-bar is expanded along the [001]-direction. 

\begin{figure}[ht]
	\includegraphics[clip=true,width=8cm]{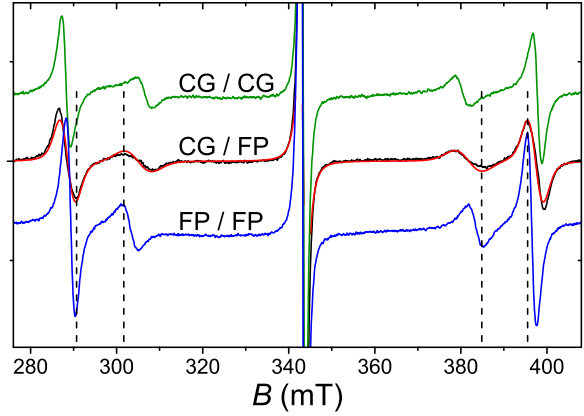}
	\caption{\label{Fig2} EPR spectra of “cubic” Fe$^{3+}$ centers in a single-crystal (001)-oriented SrTiO$_3$ plate with the combinations of the faces quality: CG/CG (green line), CG/FP (black), FP/FP (blue) measured with B$\mathbf \parallel [001]$; red line is a simulation of the CG/FP sample with a linear distribution of the axial crystal field parameter $B_2^0$ (see text). Dotted lines indicate the positions of the EPR components for cubic-symmetry center.}
\end{figure}

Studying the effects like the one we report, one should consider a residual strain due to mechanical treatments as a source of the phenomena. Therefore, we studied the dependence of the effect on the surface quality in a thin (001)-oriented STO:Fe plate. EPR spectra of the same $2 \times 2 \times 0.4$~mm$^3$ piece of STO:Fe with varied faces quality are compared in Fig.~\ref{Fig2}. Initially, both large faces of the plate were of the CG-grade roughness (CG/CG). Then, one of the large faces was carefully polished to achieve the FP-grade (FP/CG). The third spectrum was measured after fine polishing of both large faces of the plate (FP/FP). EPR spectrum was measured at each step. Note that during polishing the plate thickness decreased by less than $1\%$ of its original value (400~$\mu$m). Now, we compare the spectra of the plates with the same quality of large faces (CG/CG vs FP/FP). First, the widths of the fine structure components for these two cases is approximately identical. This indicates a substantial homogeneity of the bulk of the samples. Second, polishing of both large faces of the plate has led to a noticeable decrease in anisotropy: the fine structure of the EPR spectrum of the Fe$^{3+}$ ions became closer to the structure of the cubic center (dotted lines in Fig.~\ref{Fig2}). Axial crystal field $B_2^0$-parameter values are $+ 11.0$~MHz and $+ 4.3$~MHz for CG/CG and FP/FP surface grades, respectively. Thus, \textit{the observed effect is undoubtedly related to the quality of the faces}: the worse the quality, the greater the anisotropy. The second important conclusion is that \textit{the effect is not associated with a residual stress} accumulated during sample processing. Fine polishing represents a rough impact, accompanied by mechanical destruction of surface irregularities, thus it is hard to expect that residual deformations under this kind of treatment can decrease. However, we observed a significant reduction of the tetragonal distortion. 

In the Fe$^{3+}$-ion EPR spectrum of a (001)-plate with different quality of large faces (CG/FP), in the out-of-plane orientation (B$\mathbf\parallel[001]$) a significant broadening of the fine structure components is observed. Moreover, the resonance fields for the $\left| \pm\frac{3}{2}\right\rangle \rightarrow \left|\pm\frac{5}{2}\right\rangle $  transitions at $\sim 305$~mT and $\sim 380$~mT (most sensitive to deformations) are distributed approximately within the values for CG/CG and FP/FP plates. The width of these components in the in-plane orientation (B$\mathbf\parallel[100]$) did not experience any significant modification with polishing. This observation unambiguously reflects the the distribution of the axial deformation along the plate normal. Figure~\ref{Fig2} also presents the result of the EPR spectrum simulation for the Fe$^{3+}$ centers with a linear distribution of axial deformation between the values characteristic for CG/CG and FP/FP plates. It can be seen that such an approximation makes it possible to fairly well describe the structure of the spectrum and the shape of its components. This fact, in turn, indicates that the \textit{surface quality is a factor determining to the high extent the magnitude of the axial deformation} in the near-surface area of a plate.

\begin{figure}[ht]
	\includegraphics[clip=true,width=8cm]{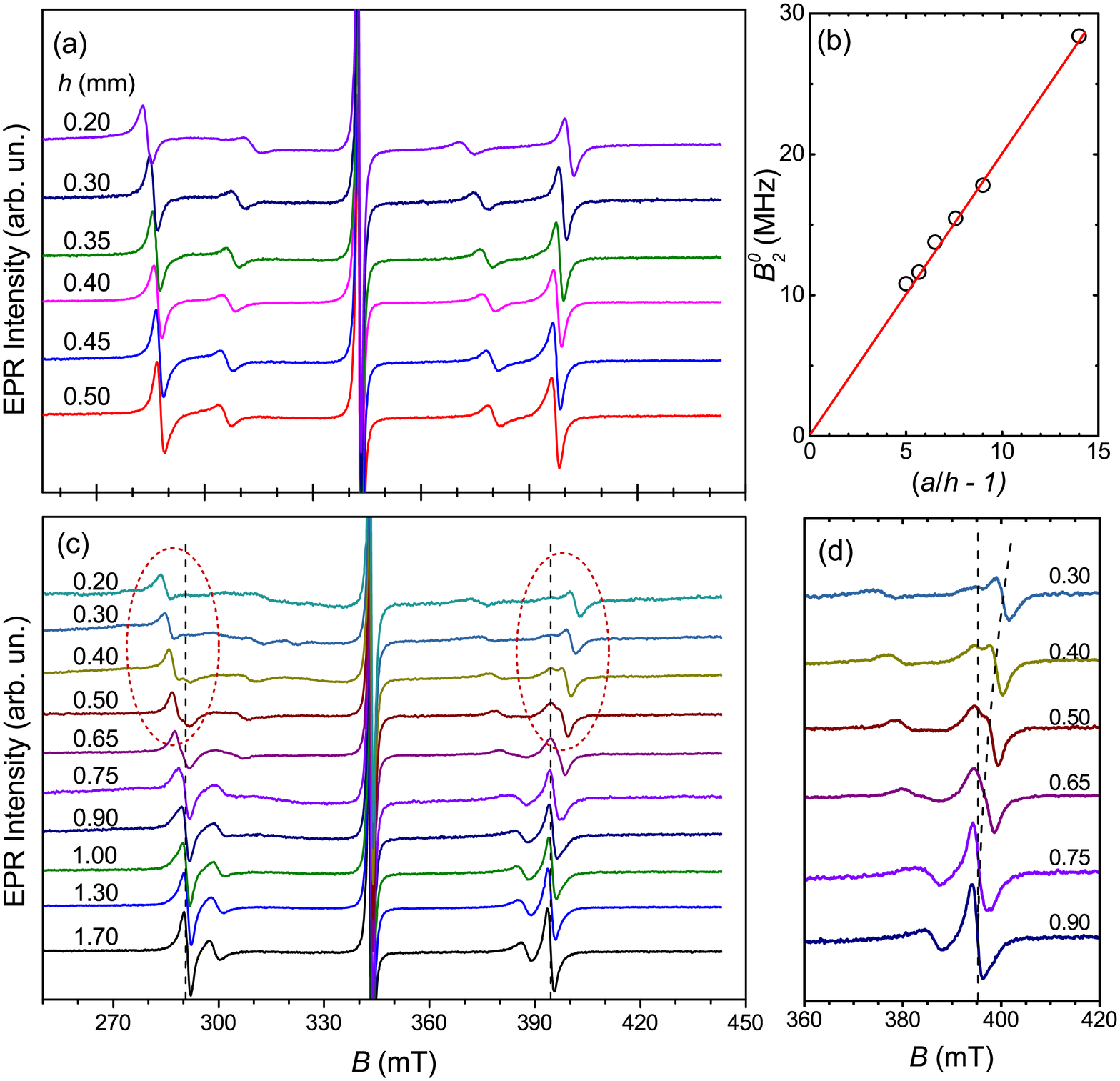}
	\caption{\label{Fig3} Dependence of the EPR spectrum of Fe$^{3+}$ ions in oriented SrTiO$_3$ single crystals: (a) for initially thin (001)-plate with a decrease in its thickness; (b) dependence of the tetragonal distortion parameter on the ratio of the base side to the plate thickness; (c) for a sample with the shape of a rectangular prism with a square base and a height initially larger than the base side; (d) magnified spectra of panel (c) with a pronounced splitting of the components. Spectra intensities are normalized to the intensity of the line corresponding to the $\left| +\frac{1}{2} \right\rangle \rightarrow \left| -\frac{1}{2} \right\rangle $ transition at $\sim 343$~mT.}
\end{figure}

To achieve a deeper insight, we studied the dependence of the axial distortion value and its distribution over the volume on a sample shape. Below we present the results for two STO:Fe samples with the right square prism shape: the first was initially a thin plate with $a = 2.5$~mm and $h_{init} = 0.50$~mm, and the second at the beginning had a height $h$ larger than a base side $a$ ($h_{init} = 1.70$~mm and $a = 1.00$~mm). All faces of both samples were of the CG-grade. In the course of this experiment, the samples’ height $h \mathbf\parallel [001]$ was reduced by grinding in a step-by-step manner; EPR spectrum was acquired at each step. The results for the plate are shown in Fig.~\ref{Fig3},~a. A clear evolution of the spectrum structure is observed. The inset shows the dependence of the axial distortion ($B_2^0$) on thickness: we find that the magnitude of the distortion grows with the plate thickness decrease and is nicely proportional to $(a/h – 1)$. Small linewidths of the fine structure components of the spectra indicate near-uniform distribution of the axial distortion over the sample volume.

The dependence of the EPR spectrum structure on the height of the second sample ($h_{init} > a$) is qualitatively different (Fig.~\ref{Fig3},~b). Here, at relatively large thicknesses ($h > 0.7$~mm), the observed EPR spectrum is characteristic for Fe$^{3+}$ centers free of the axial distortion, identical to that of the cubic Fe$^{3+}$ impurity in STO \cite{muller1958paramagnetische,unoki1967electron}. At $h < 0.7$~mm, the signals from two paramagnetic Fe$^{3+}$ centers with different local symmetry can be distinguished in the EPR spectrum. One of them is the cubic-symmetry center, another possesses the tetragonal symmetry. Axial distortion and relative intensity of the signal of the axial center increase with the decrease of the height $h$. At a thickness of $h < 0.3$~mm, the signals of the axial Fe$^{3+}$ centers dominate in the spectrum. Considering the Fe$^{3+}$ centers as paramagnetic probes, we conclude that initially the majority of the sample volume had the cubic symmetry. With a decrease of the sample height, the volume fraction with an axial symmetry first became noticeable and later took up the entire volume of the sample. It is important that in the observed spectrum the signals from the two centers with well-defined magnitude of the distortion are clearly revealed, and it is certainly not an average signal over two extreme cases. Such a set of experimental facts convincingly indicates the formation of a low (axial) symmetry phase in macroscopic, with hundreds of microns thickness, near-surface volumes with a small transient area between it and a high-symmetry (cubic) phase.

\begin{figure}[ht]
	\includegraphics[clip=true,width=8cm]{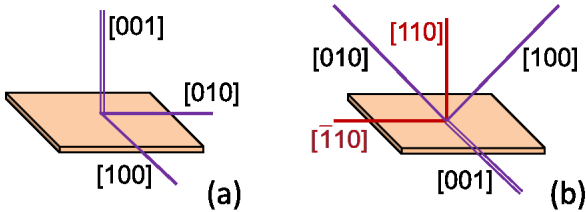}
	\caption{\label{Fig4} Crystallographic directions in (001)-oriented (a) and (110)-oriented (b) STO plates. Colors indicate structurally equivalent directions in the cubic phase of SrTiO$_3$; single and double lines show the directions in SrTiO$_3$ plates that turned out structurally inequivalent according to EPR-spectra of the impurity Fe$^{3+}$ ions.}
\end{figure}

The observed tendency of the STO crystal plates to lower the symmetry reveals a pronounced selectivity. So, for square (001)-oriented plates, the structure of the EPR spectra of Fe$^{3+}$ ions is different for B$\mathbf\parallel[001]$ and B$\mathbf\parallel[100]$, while for B$\mathbf\parallel[100]$ and B$\mathbf\parallel[010]$, the spectra are identical. Thus, in (001)-plates the directions [001] and [100] are inequivalent, and the directions [100] and [010] lying in the plane are equivalent (Fig.~\ref{Fig4}). That is, the crystal structure of the STO matrix has axial (tetragonal) symmetry with the $c$-axis parallel to the normal to the plane. Comparison of the EPR spectra of the (110)-oriented STO:Fe plate shows that crystallographically equivalent out-of-plane $[110]$ and in-plane $[\check110]$  directions are equivalent; simultaneously, the in-plane [001] direction, on the one hand, and [100] and [010], which make up an angle of 45~degrees with the plane, on the other hand, turn out to be structurally inequivalent, while [100] and [010] are equivalent (Fig.~\ref{Fig4}). This is possible for a plate with a tetragonal crystal structure with the $c$-axis along [001] lying in its plane. For the (111)-oriented STO plate, in which all the $\left\langle 100 \right\rangle $ directions are equivalent relative to the plane of the sample and the directions $\left\langle 111 \right\rangle $ are inequivalent, all the directions of the $\left\langle 100 \right\rangle $- and $\left\langle 111 \right\rangle $-types, according to the structure of the Fe$^{3+}$-ion EPR spectra, are structurally equivalent within each type. Thus, a symmetry lowering manifests itself in the samples (plates) where the four-fold cubic axes are inequivalent with respect to large faces with an imperfect surface; the symmetry is lowered to tetragonal. The SrTiO$_3$ (111)-plate retains cubic symmetry. Equivalency of the out-of-plane [110] and in-plane $\left\langle 110 \right\rangle $  directions of the (110)-plate clearly indicates that the observed effect is not associated with the residual strain accumulated during surface treatment.

A question on polar or non-polar character of the revealed tetragonal phase will be addressed by us in a separate paper. Briefly, an observation of quadratic electric-field effect for both the STO:Fe and STO:Mn (001)-oriented thin plates clearly shows that the state is non-polar.
The described set of experimental observations unambiguously indicates the structural instability in oriented STO single crystal plates with respect to a symmetry lowering from cubic to tetragonal phase, the last possessing a simple compressed (for plates) along the [001] direction structure, different from the antiferrodistortive one. This instability is inherent to the cubic phase, and the transition to a low-symmetry state is induced by the surface. 

Existence of the near-surface lattice distortions in the high-symmetry phase of crystals above the phase transition temperature was considered theoretically in Refs.~\onlinecite{levan83,kaganov71}, and for STO it was extensively studied experimentally (see Ref.~\onlinecite{dejneka2010spectroscopic} and references therein). However, STO surface distortions investigated in these works are conjugated to the AFD $O_{1h} \rightarrow D^{18}_{4h}$ transition order parameter accompanied by the lattice cell doubling and differ in symmetry from the ones we found. In addition, they are restricted to the near-surface layer that is several lattice constants thick, much smaller than in our case.

The fact that the tetragonal STO phase manifested in our experiments has a structure different from the known cubic and AFD phases leads to the assumption that it can be stabilized, at least at T $> 105$ K, due to the influence of an imperfect, specifically oriented surface. We cannot exclude some influence of crystal structural imperfections, stresses and defects that arise during grinding and polishing, however, they do not determine the effect under discussion. In our opinion, the boundary conditions, including the state of the surface, as well as the geometry and dimensions of the sample, are crucial. The proposed concept requires further investigations.

The effect we report is of fundamental, qualitative character, since in the small, less than a millimeter size samples with the shape of oriented plates or bars with imperfect surfaces, a symmetry lowering of the crystal structure to tetragonal is observed. The value of the distortion in this new tetragonal phase is of the same order as that found in the AFD-phase below 10~K. The effect described in this work can represent a significant source of the variety of phenomena observed in strontium titanate, and should be taken into account when interpreting various experimental observations. It can cause sample dependence of particular effects in STO. At the same time, it can potentially become a technologically significant degree of freedom for fine-tuning of the STO substrate structure for the synthesis of thin-films on it: (i) (001)-plates with imperfect faces are compressed along the [001] direction; (110)-plates are elongated along [001] and (ii) surface roughness and a plate thickness define the magnitude of a distortion. Importantly, $\{111\}$-oriented STO plate possesses the cubic symmetry regardless of the surface roughness.

The data that support the findings of this study are available from the corresponding author upon reasonable request.

\begin{acknowledgments}
 The work was performed with the support from the subsidy allocated to Kazan Federal University for the state assignment in the sphere of scientific activities No.~FZSM-2020-0050
 and from M\`{S}MT, Project No. SOLID21 - CZ.02.1.01/0.0/0.0/16\_019/0000760.
\end{acknowledgments}

\bibliography{literature}

\begin{thebibliography}{28}%
\makeatletter
\providecommand \@ifxundefined [1]{%
 \@ifx{#1\undefined}
}%
\providecommand \@ifnum [1]{%
 \ifnum #1\expandafter \@firstoftwo
 \else \expandafter \@secondoftwo
 \fi
}%
\providecommand \@ifx [1]{%
 \ifx #1\expandafter \@firstoftwo
 \else \expandafter \@secondoftwo
 \fi
}%
\providecommand \natexlab [1]{#1}%
\providecommand \enquote  [1]{``#1''}%
\providecommand \bibnamefont  [1]{#1}%
\providecommand \bibfnamefont [1]{#1}%
\providecommand \citenamefont [1]{#1}%
\providecommand \href@noop [0]{\@secondoftwo}%
\providecommand \href [0]{\begingroup \@sanitize@url \@href}%
\providecommand \@href[1]{\@@startlink{#1}\@@href}%
\providecommand \@@href[1]{\endgroup#1\@@endlink}%
\providecommand \@sanitize@url [0]{\catcode `\\12\catcode `\$12\catcode
  `\&12\catcode `\#12\catcode `\^12\catcode `\_12\catcode `\%12\relax}%
\providecommand \@@startlink[1]{}%
\providecommand \@@endlink[0]{}%
\providecommand \url  [0]{\begingroup\@sanitize@url \@url }%
\providecommand \@url [1]{\endgroup\@href {#1}{\urlprefix }}%
\providecommand \urlprefix  [0]{URL }%
\providecommand \Eprint [0]{\href }%
\providecommand \doibase [0]{http://dx.doi.org/}%
\providecommand \selectlanguage [0]{\@gobble}%
\providecommand \bibinfo  [0]{\@secondoftwo}%
\providecommand \bibfield  [0]{\@secondoftwo}%
\providecommand \translation [1]{[#1]}%
\providecommand \BibitemOpen [0]{}%
\providecommand \bibitemStop [0]{}%
\providecommand \bibitemNoStop [0]{.\EOS\space}%
\providecommand \EOS [0]{\spacefactor3000\relax}%
\providecommand \BibitemShut  [1]{\csname bibitem#1\endcsname}%
\let\auto@bib@innerbib\@empty
\bibitem [{\citenamefont {M{\"u}ller}\ and\ \citenamefont
  {Burkard}(1979)}]{muller1979srti}%
  \BibitemOpen
  \bibfield  {author} {\bibinfo {author} {\bibfnamefont {K.~A.}\ \bibnamefont
  {M{\"u}ller}}\ and\ \bibinfo {author} {\bibfnamefont {H.}~\bibnamefont
  {Burkard}},\ }\bibfield  {title} {\enquote {\bibinfo {title}
  {{S}r{T}i{O}$_3$: {A}n intrinsic quantum paraelectric below 4 {K}},}\
  }\href@noop {} {\bibfield  {journal} {\bibinfo  {journal} {Physical Review
  B}\ }\textbf {\bibinfo {volume} {19}},\ \bibinfo {pages} {3593} (\bibinfo
  {year} {1979})}\BibitemShut {NoStop}%
\bibitem [{\citenamefont {Barrett}(1952)}]{barrett1952dielectric}%
  \BibitemOpen
  \bibfield  {author} {\bibinfo {author} {\bibfnamefont {J.~H.}\ \bibnamefont
  {Barrett}},\ }\bibfield  {title} {\enquote {\bibinfo {title} {Dielectric
  constant in perovskite type crystals},}\ }\href@noop {} {\bibfield  {journal}
  {\bibinfo  {journal} {Physical Review}\ }\textbf {\bibinfo {volume} {86}},\
  \bibinfo {pages} {118} (\bibinfo {year} {1952})}\BibitemShut {NoStop}%
\bibitem [{\citenamefont {M{\"u}ller}(1958)}]{muller1958paramagnetische}%
  \BibitemOpen
  \bibfield  {author} {\bibinfo {author} {\bibfnamefont {K.~A.}\ \bibnamefont
  {M{\"u}ller}},\ }\emph {\bibinfo {title} {Paramagnetische {R}esonanz von
  {F}e$^{3+}$ in {S}r{T}i{O}$_3$ {E}inkristallen}},\ \href@noop {} {Ph.D.
  thesis},\ \bibinfo  {school} {ETH Zurich} (\bibinfo {year}
  {1958})\BibitemShut {NoStop}%
\bibitem [{\citenamefont {Fleury}, \citenamefont {Scott},\ and\ \citenamefont
  {Worlock}(1968)}]{fleury1968soft}%
  \BibitemOpen
  \bibfield  {author} {\bibinfo {author} {\bibfnamefont {P.}~\bibnamefont
  {Fleury}}, \bibinfo {author} {\bibfnamefont {J.}~\bibnamefont {Scott}}, \
  and\ \bibinfo {author} {\bibfnamefont {J.}~\bibnamefont {Worlock}},\
  }\bibfield  {title} {\enquote {\bibinfo {title} {Soft phonon modes and the
  110 {K} phase transition in {S}r{T}i{O}$_3$},}\ }\href@noop {} {\bibfield
  {journal} {\bibinfo  {journal} {Phys. Rev. Lett.}\ }\textbf {\bibinfo
  {volume} {21}},\ \bibinfo {pages} {16} (\bibinfo {year} {1968})}\BibitemShut
  {NoStop}%
\bibitem [{\citenamefont {Courtens}(1972)}]{courtens1972birefringence}%
  \BibitemOpen
  \bibfield  {author} {\bibinfo {author} {\bibfnamefont {E.}~\bibnamefont
  {Courtens}},\ }\bibfield  {title} {\enquote {\bibinfo {title} {Birefringence
  of {S}r{T}i{O}$_3$ produced by the 105 {K} {S}tructural {P}hase
  {T}ransition},}\ }\href@noop {} {\bibfield  {journal} {\bibinfo  {journal}
  {Phys. Rev. Lett.}\ }\textbf {\bibinfo {volume} {29}},\ \bibinfo {pages}
  {1380} (\bibinfo {year} {1972})}\BibitemShut {NoStop}%
\bibitem [{\citenamefont {Cowley}(1996)}]{cowley1996phase}%
  \BibitemOpen
  \bibfield  {author} {\bibinfo {author} {\bibfnamefont {R.~A.}\ \bibnamefont
  {Cowley}},\ }\bibfield  {title} {\enquote {\bibinfo {title} {The phase
  transition of strontium titanate},}\ }\href@noop {} {\bibfield  {journal}
  {\bibinfo  {journal} {Philosophical Transactions of the Royal Society of
  London. Series A: Mathematical, Physical and Engineering Sciences}\ }\textbf
  {\bibinfo {volume} {354}},\ \bibinfo {pages} {2799--2814} (\bibinfo {year}
  {1996})}\BibitemShut {NoStop}%
\bibitem [{\citenamefont {Yamanaka}\ \emph {et~al.}(2000)\citenamefont
  {Yamanaka}, \citenamefont {Kataoka}, \citenamefont {Inaba}, \citenamefont
  {Inoue}, \citenamefont {Hehlen},\ and\ \citenamefont
  {Courtens}}]{yamanaka2000evidence}%
  \BibitemOpen
  \bibfield  {author} {\bibinfo {author} {\bibfnamefont {A.}~\bibnamefont
  {Yamanaka}}, \bibinfo {author} {\bibfnamefont {M.}~\bibnamefont {Kataoka}},
  \bibinfo {author} {\bibfnamefont {Y.}~\bibnamefont {Inaba}}, \bibinfo
  {author} {\bibfnamefont {K.}~\bibnamefont {Inoue}}, \bibinfo {author}
  {\bibfnamefont {B.}~\bibnamefont {Hehlen}}, \ and\ \bibinfo {author}
  {\bibfnamefont {E.}~\bibnamefont {Courtens}},\ }\bibfield  {title} {\enquote
  {\bibinfo {title} {Evidence for competing orderings in strontium titanate
  from hyper-raman scattering spectroscopy},}\ }\href@noop {} {\bibfield
  {journal} {\bibinfo  {journal} {EPL (Europhysics Letters)}\ }\textbf
  {\bibinfo {volume} {50}},\ \bibinfo {pages} {688} (\bibinfo {year}
  {2000})}\BibitemShut {NoStop}%
\bibitem [{\citenamefont {Kvyatkovskii}(2001)}]{Kvyat2001}%
  \BibitemOpen
  \bibfield  {author} {\bibinfo {author} {\bibfnamefont {O.}~\bibnamefont
  {Kvyatkovskii}},\ }\bibfield  {title} {\enquote {\bibinfo {title} {Quantum
  effects in incipient and low-temperature ferroelectrics},}\ }\href@noop {}
  {\bibfield  {journal} {\bibinfo  {journal} {Phys. Solid State}\ }\textbf
  {\bibinfo {volume} {43}},\ \bibinfo {pages} {1401} (\bibinfo {year}
  {2001})}\BibitemShut {NoStop}%
\bibitem [{\citenamefont {Lemanov}, \citenamefont {Smirnova},\ and\
  \citenamefont {Tarakanov}(1997)}]{Lemanov97_STO_Pb}%
  \BibitemOpen
  \bibfield  {author} {\bibinfo {author} {\bibfnamefont {V.~V.}\ \bibnamefont
  {Lemanov}}, \bibinfo {author} {\bibfnamefont {E.~P.}\ \bibnamefont
  {Smirnova}}, \ and\ \bibinfo {author} {\bibfnamefont {E.~A.}\ \bibnamefont
  {Tarakanov}},\ }\bibfield  {title} {\enquote {\bibinfo {title}
  {Ferroelectricity in {S}r{T}i{O}{$_3$}:{P}b},}\ }\href@noop {} {\bibfield
  {journal} {\bibinfo  {journal} {Ferroelectrics Lett. Sec.}\ }\textbf
  {\bibinfo {volume} {22}},\ \bibinfo {pages} {69} (\bibinfo {year}
  {1997})}\BibitemShut {NoStop}%
\bibitem [{\citenamefont {Taniuchi}\ \emph {et~al.}(2016)\citenamefont
  {Taniuchi}, \citenamefont {Motoyui}, \citenamefont {Morozumi}, \citenamefont
  {R\"{o}del}, \citenamefont {Fortuna}, \citenamefont {Santander-Syro},\ and\
  \citenamefont {Shin}}]{Taniuchi2016}%
  \BibitemOpen
  \bibfield  {author} {\bibinfo {author} {\bibfnamefont {T.}~\bibnamefont
  {Taniuchi}}, \bibinfo {author} {\bibfnamefont {Y.}~\bibnamefont {Motoyui}},
  \bibinfo {author} {\bibfnamefont {K.}~\bibnamefont {Morozumi}}, \bibinfo
  {author} {\bibfnamefont {T.}~\bibnamefont {R\"{o}del}}, \bibinfo {author}
  {\bibfnamefont {F.}~\bibnamefont {Fortuna}}, \bibinfo {author} {\bibfnamefont
  {A.}~\bibnamefont {Santander-Syro}}, \ and\ \bibinfo {author} {\bibfnamefont
  {S.}~\bibnamefont {Shin}},\ }\bibfield  {title} {\enquote {\bibinfo {title}
  {Imaging of room-temperature ferromagnetic nano-domains at the surface of a
  non-magnetic oxide},}\ }\href@noop {} {\bibfield  {journal} {\bibinfo
  {journal} {Nature Commun.}\ }\textbf {\bibinfo {volume} {7}},\ \bibinfo
  {pages} {1} (\bibinfo {year} {2016})}\BibitemShut {NoStop}%
\bibitem [{\citenamefont {Tufte}\ and\ \citenamefont
  {Chapman}(1967)}]{Tufte67}%
  \BibitemOpen
  \bibfield  {author} {\bibinfo {author} {\bibfnamefont {O.~N.}\ \bibnamefont
  {Tufte}}\ and\ \bibinfo {author} {\bibfnamefont {P.~W.}\ \bibnamefont
  {Chapman}},\ }\bibfield  {title} {\enquote {\bibinfo {title} {Electron
  mobility in semiconducting strontium titanate},}\ }\href {\doibase
  10.1103/PhysRev.155.796} {\bibfield  {journal} {\bibinfo  {journal} {Phys.
  Rev.}\ }\textbf {\bibinfo {volume} {155}},\ \bibinfo {pages} {796--802}
  (\bibinfo {year} {1967})}\BibitemShut {NoStop}%
\bibitem [{\citenamefont {Schooley}, \citenamefont {Hosler},\ and\
  \citenamefont {Cohen}(1964)}]{schooley1964superconductivity}%
  \BibitemOpen
  \bibfield  {author} {\bibinfo {author} {\bibfnamefont {J.}~\bibnamefont
  {Schooley}}, \bibinfo {author} {\bibfnamefont {W.}~\bibnamefont {Hosler}}, \
  and\ \bibinfo {author} {\bibfnamefont {M.~L.}\ \bibnamefont {Cohen}},\
  }\bibfield  {title} {\enquote {\bibinfo {title} {Superconductivity in
  semiconducting {S}r{T}i{O}$_3$},}\ }\href@noop {} {\bibfield  {journal}
  {\bibinfo  {journal} {Phys. Rev. Lett.}\ }\textbf {\bibinfo {volume} {12}},\
  \bibinfo {pages} {474} (\bibinfo {year} {1964})}\BibitemShut {NoStop}%
\bibitem [{\citenamefont {Schooley}\ \emph {et~al.}(1965)\citenamefont
  {Schooley}, \citenamefont {Hosler}, \citenamefont {Ambler}, \citenamefont
  {Becker}, \citenamefont {Cohen},\ and\ \citenamefont
  {Koonce}}]{schooley1965dependence}%
  \BibitemOpen
  \bibfield  {author} {\bibinfo {author} {\bibfnamefont {J.}~\bibnamefont
  {Schooley}}, \bibinfo {author} {\bibfnamefont {W.}~\bibnamefont {Hosler}},
  \bibinfo {author} {\bibfnamefont {E.}~\bibnamefont {Ambler}}, \bibinfo
  {author} {\bibfnamefont {J.}~\bibnamefont {Becker}}, \bibinfo {author}
  {\bibfnamefont {M.~L.}\ \bibnamefont {Cohen}}, \ and\ \bibinfo {author}
  {\bibfnamefont {C.}~\bibnamefont {Koonce}},\ }\bibfield  {title} {\enquote
  {\bibinfo {title} {Dependence of the superconducting transition temperature
  on carrier concentration in semiconducting {S}r{T}i{O}$_3$},}\ }\href@noop {}
  {\bibfield  {journal} {\bibinfo  {journal} {Physical Review Letters}\
  }\textbf {\bibinfo {volume} {14}},\ \bibinfo {pages} {305} (\bibinfo {year}
  {1965})}\BibitemShut {NoStop}%
\bibitem [{\citenamefont {Karg}\ \emph {et~al.}(2006)\citenamefont {Karg},
  \citenamefont {Meijer}, \citenamefont {Widmer},\ and\ \citenamefont
  {Bednorz}}]{karg2006electrical}%
  \BibitemOpen
  \bibfield  {author} {\bibinfo {author} {\bibfnamefont {S.}~\bibnamefont
  {Karg}}, \bibinfo {author} {\bibfnamefont {G.}~\bibnamefont {Meijer}},
  \bibinfo {author} {\bibfnamefont {D.}~\bibnamefont {Widmer}}, \ and\ \bibinfo
  {author} {\bibfnamefont {J.}~\bibnamefont {Bednorz}},\ }\bibfield  {title}
  {\enquote {\bibinfo {title} {Electrical-stress-induced conductivity increase
  in {S}r{T}i{O}$_3$ films},}\ }\href@noop {} {\bibfield  {journal} {\bibinfo
  {journal} {Applied Physics Letters}\ }\textbf {\bibinfo {volume} {89}},\
  \bibinfo {pages} {072106} (\bibinfo {year} {2006})}\BibitemShut {NoStop}%
\bibitem [{\citenamefont {Szot}\ \emph {et~al.}(2006)\citenamefont {Szot},
  \citenamefont {Speier}, \citenamefont {Bihlmayer},\ and\ \citenamefont
  {Waser}}]{szot2006switching}%
  \BibitemOpen
  \bibfield  {author} {\bibinfo {author} {\bibfnamefont {K.}~\bibnamefont
  {Szot}}, \bibinfo {author} {\bibfnamefont {W.}~\bibnamefont {Speier}},
  \bibinfo {author} {\bibfnamefont {G.}~\bibnamefont {Bihlmayer}}, \ and\
  \bibinfo {author} {\bibfnamefont {R.}~\bibnamefont {Waser}},\ }\bibfield
  {title} {\enquote {\bibinfo {title} {Switching the electrical resistance of
  individual dislocations in single-crystalline {S}r{T}i{O}$_3$},}\ }\href@noop
  {} {\bibfield  {journal} {\bibinfo  {journal} {Nature Materials}\ }\textbf
  {\bibinfo {volume} {5}},\ \bibinfo {pages} {312--320} (\bibinfo {year}
  {2006})}\BibitemShut {NoStop}%
\bibitem [{\citenamefont {Makarova}\ \emph {et~al.}(2010)\citenamefont
  {Makarova}, \citenamefont {Dejneka}, \citenamefont {Franc}, \citenamefont
  {Drahokoupil}, \citenamefont {Jastrabik},\ and\ \citenamefont
  {Trepakov}}]{Makarova2010}%
  \BibitemOpen
  \bibfield  {author} {\bibinfo {author} {\bibfnamefont {M.}~\bibnamefont
  {Makarova}}, \bibinfo {author} {\bibfnamefont {A.}~\bibnamefont {Dejneka}},
  \bibinfo {author} {\bibfnamefont {J.}~\bibnamefont {Franc}}, \bibinfo
  {author} {\bibfnamefont {J.}~\bibnamefont {Drahokoupil}}, \bibinfo {author}
  {\bibfnamefont {L.}~\bibnamefont {Jastrabik}}, \ and\ \bibinfo {author}
  {\bibfnamefont {V.}~\bibnamefont {Trepakov}},\ }\bibfield  {title} {\enquote
  {\bibinfo {title} {Soft chemistry preparation methods and properties of
  strontium titanate nanoparticles},}\ }\href@noop {} {\bibfield  {journal}
  {\bibinfo  {journal} {Optical Materials}\ }\textbf {\bibinfo {volume} {32}},\
  \bibinfo {pages} {803} (\bibinfo {year} {2010})}\BibitemShut {NoStop}%
\bibitem [{\citenamefont {Tyunina}\ \emph {et~al.}(2009)\citenamefont
  {Tyunina}, \citenamefont {Narkilahti}, \citenamefont {Levoska}, \citenamefont
  {Chvostova}, \citenamefont {Dejneka}, \citenamefont {Trepakov},\ and\
  \citenamefont {Zelezny}}]{Tyunina2009}%
  \BibitemOpen
  \bibfield  {author} {\bibinfo {author} {\bibfnamefont {M.}~\bibnamefont
  {Tyunina}}, \bibinfo {author} {\bibfnamefont {J.}~\bibnamefont {Narkilahti}},
  \bibinfo {author} {\bibfnamefont {J.}~\bibnamefont {Levoska}}, \bibinfo
  {author} {\bibfnamefont {D.}~\bibnamefont {Chvostova}}, \bibinfo {author}
  {\bibfnamefont {A.}~\bibnamefont {Dejneka}}, \bibinfo {author} {\bibfnamefont
  {V.}~\bibnamefont {Trepakov}}, \ and\ \bibinfo {author} {\bibfnamefont
  {V.}~\bibnamefont {Zelezny}},\ }\bibfield  {title} {\enquote {\bibinfo
  {title} {Ultrathin {S}r{T}i{O}{$_3$} films: epitaxy and optical
  properties},}\ }\href@noop {} {\bibfield  {journal} {\bibinfo  {journal} {J.
  Phys. Cond. Matt.}\ }\textbf {\bibinfo {volume} {21}},\ \bibinfo {pages}
  {232203} (\bibinfo {year} {2009})}\BibitemShut {NoStop}%
\bibitem [{\citenamefont {Alexandrov}\ \emph {et~al.}(2009)\citenamefont
  {Alexandrov}, \citenamefont {Kotomin}, \citenamefont {Maier},\ and\
  \citenamefont {Evarestov}}]{alexandrov2009first}%
  \BibitemOpen
  \bibfield  {author} {\bibinfo {author} {\bibfnamefont {V.}~\bibnamefont
  {Alexandrov}}, \bibinfo {author} {\bibfnamefont {E.}~\bibnamefont {Kotomin}},
  \bibinfo {author} {\bibfnamefont {J.}~\bibnamefont {Maier}}, \ and\ \bibinfo
  {author} {\bibfnamefont {R.}~\bibnamefont {Evarestov}},\ }\bibfield  {title}
  {\enquote {\bibinfo {title} {First-principles study of bulk and surface
  oxygen vacancies in {S}r{T}i{O}$_3$ crystal},}\ }\href@noop {} {\bibfield
  {journal} {\bibinfo  {journal} {The European Physical Journal B}\ }\textbf
  {\bibinfo {volume} {72}},\ \bibinfo {pages} {53--57} (\bibinfo {year}
  {2009})}\BibitemShut {NoStop}%
\bibitem [{\citenamefont {Oja}\ \emph {et~al.}(2012)\citenamefont {Oja},
  \citenamefont {Tyunina}, \citenamefont {Yao}, \citenamefont {Pinomaa},
  \citenamefont {Kocourek}, \citenamefont {Dejneka}, \citenamefont {Stupakov},
  \citenamefont {Jelinek}, \citenamefont {Trepakov}, \citenamefont {van
  Dijken},\ and\ \citenamefont {Nieminen}}]{oja2012d}%
  \BibitemOpen
  \bibfield  {author} {\bibinfo {author} {\bibfnamefont {R.}~\bibnamefont
  {Oja}}, \bibinfo {author} {\bibfnamefont {M.}~\bibnamefont {Tyunina}},
  \bibinfo {author} {\bibfnamefont {L.}~\bibnamefont {Yao}}, \bibinfo {author}
  {\bibfnamefont {T.}~\bibnamefont {Pinomaa}}, \bibinfo {author} {\bibfnamefont
  {T.}~\bibnamefont {Kocourek}}, \bibinfo {author} {\bibfnamefont
  {A.}~\bibnamefont {Dejneka}}, \bibinfo {author} {\bibfnamefont
  {O.}~\bibnamefont {Stupakov}}, \bibinfo {author} {\bibfnamefont
  {M.}~\bibnamefont {Jelinek}}, \bibinfo {author} {\bibfnamefont
  {V.}~\bibnamefont {Trepakov}}, \bibinfo {author} {\bibfnamefont
  {S.}~\bibnamefont {van Dijken}}, \ and\ \bibinfo {author} {\bibfnamefont
  {R.~M.}\ \bibnamefont {Nieminen}},\ }\bibfield  {title} {\enquote {\bibinfo
  {title} {$d^0$ ferromagnetic interface between nonmagnetic perovskites},}\
  }\href@noop {} {\bibfield  {journal} {\bibinfo  {journal} {Physical Review
  Letters}\ }\textbf {\bibinfo {volume} {109}},\ \bibinfo {pages} {127207}
  (\bibinfo {year} {2012})}\BibitemShut {NoStop}%
\bibitem [{\citenamefont {Unoki}\ and\ \citenamefont
  {Sakudo}(1967)}]{unoki1967electron}%
  \BibitemOpen
  \bibfield  {author} {\bibinfo {author} {\bibfnamefont {H.}~\bibnamefont
  {Unoki}}\ and\ \bibinfo {author} {\bibfnamefont {T.}~\bibnamefont {Sakudo}},\
  }\bibfield  {title} {\enquote {\bibinfo {title} {Electron spin resonance of
  {F}e$^{3+}$ in {S}r{T}i{O}$_3$ with special reference to the 110 {K} phase
  transition},}\ }\href@noop {} {\bibfield  {journal} {\bibinfo  {journal}
  {Journal of the Physical Society of Japan}\ }\textbf {\bibinfo {volume}
  {23}},\ \bibinfo {pages} {546--552} (\bibinfo {year} {1967})}\BibitemShut
  {NoStop}%
\bibitem [{\citenamefont {Rimai}\ and\ \citenamefont
  {DeMars}(1962)}]{rimai1962electron}%
  \BibitemOpen
  \bibfield  {author} {\bibinfo {author} {\bibfnamefont {L.}~\bibnamefont
  {Rimai}}\ and\ \bibinfo {author} {\bibfnamefont {G.}~\bibnamefont {DeMars}},\
  }\bibfield  {title} {\enquote {\bibinfo {title} {Electron paramagnetic
  resonance of trivalent gadolinium ions in strontium and barium titanates},}\
  }\href@noop {} {\bibfield  {journal} {\bibinfo  {journal} {Physical Review}\
  }\textbf {\bibinfo {volume} {127}},\ \bibinfo {pages} {702} (\bibinfo {year}
  {1962})}\BibitemShut {NoStop}%
\bibitem [{\citenamefont {Stoll}\ and\ \citenamefont
  {Schweiger}(2006)}]{stoll2006easyspin}%
  \BibitemOpen
  \bibfield  {author} {\bibinfo {author} {\bibfnamefont {S.}~\bibnamefont
  {Stoll}}\ and\ \bibinfo {author} {\bibfnamefont {A.}~\bibnamefont
  {Schweiger}},\ }\bibfield  {title} {\enquote {\bibinfo {title} {Easyspin, a
  comprehensive software package for spectral simulation and analysis in
  {EPR}},}\ }\href@noop {} {\bibfield  {journal} {\bibinfo  {journal} {Journal
  of magnetic resonance}\ }\textbf {\bibinfo {volume} {178}},\ \bibinfo {pages}
  {42--55} (\bibinfo {year} {2006})}\BibitemShut {NoStop}%
\bibitem [{\citenamefont {Abraham}\ and\ \citenamefont
  {Bleaney}(1970)}]{abraham1970electron}%
  \BibitemOpen
  \bibfield  {author} {\bibinfo {author} {\bibfnamefont {A.}~\bibnamefont
  {Abraham}}\ and\ \bibinfo {author} {\bibfnamefont {B.}~\bibnamefont
  {Bleaney}},\ }\bibfield  {title} {\enquote {\bibinfo {title} {Electron
  paramagnetic resonance of transition ions},}\ }\href@noop {} {\bibfield
  {journal} {\bibinfo  {journal} {Clarendon, Oxford}\ } (\bibinfo {year}
  {1970})}\BibitemShut {NoStop}%
\bibitem [{\citenamefont {Gabbasov}\ \emph {et~al.}(2018)\citenamefont
  {Gabbasov}, \citenamefont {Gracheva}, \citenamefont {Nikitin}, \citenamefont
  {Zverev}, \citenamefont {Dejneka}, \citenamefont {Trepakov},\ and\
  \citenamefont {Yusupov}}]{gabbasov2018experimental}%
  \BibitemOpen
  \bibfield  {author} {\bibinfo {author} {\bibfnamefont {B.}~\bibnamefont
  {Gabbasov}}, \bibinfo {author} {\bibfnamefont {I.}~\bibnamefont {Gracheva}},
  \bibinfo {author} {\bibfnamefont {S.}~\bibnamefont {Nikitin}}, \bibinfo
  {author} {\bibfnamefont {D.}~\bibnamefont {Zverev}}, \bibinfo {author}
  {\bibfnamefont {A.}~\bibnamefont {Dejneka}}, \bibinfo {author} {\bibfnamefont
  {V.}~\bibnamefont {Trepakov}}, \ and\ \bibinfo {author} {\bibfnamefont
  {R.}~\bibnamefont {Yusupov}},\ }\bibfield  {title} {\enquote {\bibinfo
  {title} {Experimental evidences of the shape-induced structural distortion of
  {S}r{T}i{O}$_3$ single crystals from impurity {M}n$^{4+}$ ions electron
  paramagnetic resonance},}\ }\href@noop {} {\bibfield  {journal} {\bibinfo
  {journal} {Magnetic Resonance in Solids}\ }\textbf {\bibinfo {volume} {20}}
  (\bibinfo {year} {2018})}\BibitemShut {NoStop}%
\bibitem [{\citenamefont {Dobrov}, \citenamefont {Vieth},\ and\ \citenamefont
  {Browne}(1959)}]{dobrov1959electron}%
  \BibitemOpen
  \bibfield  {author} {\bibinfo {author} {\bibfnamefont {W.}~\bibnamefont
  {Dobrov}}, \bibinfo {author} {\bibfnamefont {R.}~\bibnamefont {Vieth}}, \
  and\ \bibinfo {author} {\bibfnamefont {M.}~\bibnamefont {Browne}},\
  }\bibfield  {title} {\enquote {\bibinfo {title} {Electron {P}aramagnetic
  {R}esonance in {S}r{T}i{O}$_3$},}\ }\href@noop {} {\bibfield  {journal}
  {\bibinfo  {journal} {Physical Review}\ }\textbf {\bibinfo {volume} {115}},\
  \bibinfo {pages} {79} (\bibinfo {year} {1959})}\BibitemShut {NoStop}%
\bibitem [{\citenamefont {Levanyuk}\ and\ \citenamefont
  {Minyukov}(1983)}]{levan83}%
  \BibitemOpen
  \bibfield  {author} {\bibinfo {author} {\bibfnamefont {A.~P.}\ \bibnamefont
  {Levanyuk}}\ and\ \bibinfo {author} {\bibfnamefont {C.~A.}\ \bibnamefont
  {Minyukov}},\ }\bibfield  {title} {\enquote {\bibinfo {title} {Evolution of
  crystal structure distortions on the surface in the vicinity of structural
  phase transitions},}\ }\href@noop {} {\bibfield  {journal} {\bibinfo
  {journal} {Physics Solid State}\ }\textbf {\bibinfo {volume} {25}},\ \bibinfo
  {pages} {2617} (\bibinfo {year} {1983})}\BibitemShut {NoStop}%
\bibitem [{\citenamefont {Kaganov}\ and\ \citenamefont
  {Omel'yanchuk}(1971)}]{kaganov71}%
  \BibitemOpen
  \bibfield  {author} {\bibinfo {author} {\bibfnamefont {M.~I.}\ \bibnamefont
  {Kaganov}}\ and\ \bibinfo {author} {\bibfnamefont {A.~M.}\ \bibnamefont
  {Omel'yanchuk}},\ }\bibfield  {title} {\enquote {\bibinfo {title}
  {Phenomenological theory of phase transition in a thin ferromagnetic
  plate},}\ }\href@noop {} {\bibfield  {journal} {\bibinfo  {journal} {Sov.
  Phys. JETP}\ }\textbf {\bibinfo {volume} {34}},\ \bibinfo {pages} {895}
  (\bibinfo {year} {1971})}\BibitemShut {NoStop}%
\bibitem [{\citenamefont {Dejneka}, \citenamefont {Trepakov},\ and\
  \citenamefont {Jastrabik}(2010)}]{dejneka2010spectroscopic}%
  \BibitemOpen
  \bibfield  {author} {\bibinfo {author} {\bibfnamefont {A.}~\bibnamefont
  {Dejneka}}, \bibinfo {author} {\bibfnamefont {V.}~\bibnamefont {Trepakov}}, \
  and\ \bibinfo {author} {\bibfnamefont {L.}~\bibnamefont {Jastrabik}},\
  }\bibfield  {title} {\enquote {\bibinfo {title} {Spectroscopic ellipsometry
  of {S}r{T}i{O}$_3$ crystals applied to antiferrodistortive surface phase
  transition},}\ }\href@noop {} {\bibfield  {journal} {\bibinfo  {journal}
  {Physica Status Solidi (b)}\ }\textbf {\bibinfo {volume} {247}},\ \bibinfo
  {pages} {1951--1955} (\bibinfo {year} {2010})}\BibitemShut {NoStop}%
\end{thebibliography}%

\end{document}